\documentstyle[12pt]{article}
\begin{document}
\begin{large}

\leftline{\hskip 12 truecm IJS.TP.99/17} 
\leftline{\hskip 12 truecm NBI-HE-99-35} 
\leftline{\hskip 11.5 truecm CERN-TH/99-288} 

\vspace{5mm}

\centerline{\bf SPINS AND CHARGES IN GRASSMANN SPACE and}

\centerline{\bf K\" AHLER SPINORS IN SPACE OF DIFFERENTIAL FORMS  }
\end{large}

\vspace{0.8cm}
\centerline{\rm NORMA MANKO\v C BOR\v STNIK\footnote[1]{The
invited talk, presented on International Workshop on Lorentz
group, CPT and Neutrinos, Zacatecas, 23-26 June, 1999.}}
\centerline{\it Dept. of Physics, University of
Ljubljana, Jadranska 19, and }
\centerline{\it J. Stefan Institute, Jamova 39,
Ljubljana, 1111, Slovenia}
\centerline{\it E-mail: norma.s.mankoc@ijs.si }

\vspace{2mm}
\centerline{\rm and}
\centerline{\rm HOLGER BECH NIELSEN}
\centerline{\it Dept. of Physics,  Niels Bohr Institute,}
\centerline{\it Blegdamsvej 17, Copenhagen, DK-2100 and}
\centerline{\it Theoretical Physics Division, CERN, CH-1211
Geneva 23}
\centerline{\it E-mail: hbech@nbvims.nbi.dk}

\abstract{ One of us got spins and charges of not
only scalars and vectors but also of spinors out of fields,
which are antisymmetric tensor fields. K\" ahler got spins of
spinors out of differential forms, which again are antisymmetric
tensor fields. Using our simple Grassmann formulation of spins
and charges of either spinors or vectors and comparing it to the
Dirac-K\" ahler formulation of spinors, we generalize the
Dirac-K\" ahler approach to vector internal degrees of freedom
and to charges of either spinors or vectors and tenzors and
point out how at all spinors can appear in both approaches.}  

\section{  Introduction.}

K\" ahler\cite{kah} has shown how to use differential forms to
describe the spin of fermions. One of us\cite{norma1} has shown
how a space of anticommuting coordinates can be used to describe
spins and charges of not only fermions but also of bosons,
unifying spins and charges for either fermions or for bosons.\\
In the present talk we point out the analogy and nice relations
between the two different  ways of achieving the  appearance of 
spin one half degrees of freedom when starting from 
pure vectors and tensors. We comment the necessity of appearance
of four copies of Dirac fermions in both approaches. \\
Comparing carefully the two approaches we generalize the K\"
ahler approach to describe also integer spins as well as charges
for either spinors or vectors, unifying spins and charges.

\section{Dirac equations in Grassmann space.}

What we call quantum mechanics in Grassmann space\cite{norma1}
is the model for going beyond the Standard Model with extra
dimensions of ordinary and anticommuting coordinates,
describing spins and charges of either fermions or bosons in an
unique way. \\
In a $d$-dimensional space-time the internal degrees of freedom
of either spinors or vectors and scalars come from the odd
Grassmannian variables $\theta^a, \quad a \in \{ 0,1,2,3,5,
\cdot ,d \}$.  \\
We write wave functions describing either spinors or vectors 
in  the form
\begin{equation}
\Phi (\theta^a) = \sum_{i=0,1,..,3,5,..,d} \quad \sum_{\{ a_1<
a_2<...<a_i\}\in \{0,1,..,3,5,..,d\} }
\alpha_{a_1, a_2,...,a_i}\theta^{a_1} 
\theta^{a_2} \cdots \theta^{a_i},
\label{phi}
\end{equation} 
where the coefficients $ \alpha_{a_1, a_2,...,a_i}$ depend on
 commuting coordinates $ x^a, \; a \in \{0,1,2,3,5,..,d \}. $
The wave function space spanned over Grassmannian coordinate
space has the dimension $2^d$. Completely analogously to usual
quantum mechanics we have the operator for the conjugate
variable $\theta^a$ to be 
\begin{equation}
p^{\theta}_a =-i\overrightarrow{\partial}_a.
\label{pt}
\end{equation} 
The rihgt arrow tells, that the derivation has to be performed
from the left hand side. These operators then obey the odd
Heisenberg algebra, which written by means of the generalized
commutators 
\begin{equation}
\{ A, B \}: = AB - ( -1)^{n_{AB}} BA,
\label{gc}
\end{equation}
where 
\begin{equation}
n_{AB}=\left\{ \begin{array}{rl} +1, \quad \hbox{if A and B have
Grassmann odd character}\\ 0, \quad \hbox{otherwise,}\end{array} \right.
\label{gen}
\end{equation}   
takes the form
\begin{equation}
\{p^{\theta a},p^{\theta b}\} = 0 = \{\theta^a, \theta^b \} , \quad
\{ p^{\theta a}, \theta^b \} = -i \eta^{ab}.
\label{pptt}
\end{equation}
Here $\eta^{ab}$ is the flat metric $\eta = diag\{1,-1,-1,...\}$.
\\
We may  define the operators 
\begin{equation}
\tilde{a}^a := i(p^{\theta a}-i\theta^a), \quad\tilde{
\tilde{a}}{ }^a :=
-(p^{\theta a}+i\theta^a),
\label{eq6}
\end{equation}
for which we can show that the $\tilde{a}^a$'s among themselves 
fulfill the Clifford 
algebra as do also the $\tilde{\tilde{a}}{ }^a$'s, while they
mutually anticommute:
\begin{equation}
\{ \tilde{a}^a, \tilde{a}^b\} = 2\eta^{ab} = \{\tilde{\tilde{a}}{
}^a, 
\tilde{\tilde{a}}{ }^b\}, \quad \{\tilde
{a}^a, \tilde{\tilde{a}}{ }^b\} = 0.
\label{eq7}
\end{equation}

\noindent
We could recognize formally
\begin{equation}
{\rm either} \quad \tilde{a}^a p_a|\Phi> = 0, \qquad {\rm or} \quad 
\tilde{\tilde{a}}{ }^a p_a|\Phi> = 0 
\label{d}
\end{equation}
as the Dirac-like equation, because of the 
above generalized
commutation relations.
Applying either the operator $\tilde{a}^a p_a$ or $\tilde{\tilde{a}}{
}^a p_a$ on the two equations we get the Klein-Gordon equation
$p^ap_a|\Phi> = 0$, where 
we define  $p_a = i\frac{\partial}{\partial x^a}$.
\\
One can, however, check that none of the two equations
(\ref{d}) have solutions which would transform as spinors with
respect to the the generators for the Lorentz transformations,
when taken in analogy with the generators of the Lorentz
transformations in ordinary space ($L^{ab} = x^a p^b - x^b p^a $)
\begin{equation}
S^{ab}:= \theta^a p^{\theta b} - \theta^b p^{\theta a}.
\label{vecs}
\end{equation}
We can write, however, these generators as the sum
\begin{equation}
S^{ab} = \tilde{S}^{ab} + \tilde{\tilde{S}}{ }^{ab},
\quad \tilde{S}^{ab} := -\frac{i}{4} [\tilde{a}^a, \tilde{a}^b], \quad
\tilde{\tilde{S}}{ }^{ab} := -\frac{i}{4}[\tilde{\tilde{a}}{
}^a,\tilde{\tilde{a}}{ }^b],  
\label{vecsp}
\end{equation}
whith $[A,B]:=AB-BA$ and recognize that the solutions of the two
equations (\ref{d}) now transform as spinors with respect to
either $ \tilde{S}^{ab}$ or $ \tilde{\tilde{S}}{ }^{ab}.$
\\
One also can easily see that the untilded, the single tilded and
the double tilded  
$S^{ab}$ obey the $d$-dimensional Lorentz generator algebra
$\{ M^{ab}, M^{cd} \} = -i(M^{ad} \eta^{bc} + M^{bc} \eta^{ad} 
- M^{ac} \eta^{bd} - M^{bd} \eta^{ac})$, when inserted for
$M^{ab}$. 
\\
We shall present this approach in more details in
section 4 when pointing out the similarities between this approach
and the K\" ahler approach and generalizing the K\" ahler
approach. 

\section{K\" ahler formulation of spinors. }

K\" ahler formulates\cite{kah} spinors in terms of wave
functions which are superpositions of the p-forms in the $d=4$ -
dimensional 
space. The 0-forms are scalars, the 1-forms are defined as dual
vectors to the (local) tangent spaces, the higher p-forms are
defined as antisymmetrized cartesian (exterior) products of the
one-form spaces. A general linear combination of forms is then
written 
\begin{equation}
u=u_0 + u_1 +...+u_d, \quad 
u_p =  \sum_{ i_1<i_2...<i_p}  a_{i_1i_2...i_p}\;
dx^{i_1}\wedge dx^{i_2}\wedge dx^{i_3} \wedge \cdots \wedge
dx^{i_p}.
\label{form} 
\end{equation}

\noindent
One can define the exterior product $\wedge$ and the Clifford
product $\vee$ among the forms. The exterior product has the
property of making the product of a p-form and a q-form to be a
(p+q)-form, if a p-form and a q-form have no common
differentials. The Clifford product $dx^a \vee$ on a p-form is
either a $p + 1$ form, if a p-form does not include a one form
$dx^a$, or a $p-1$ form, if a one form $dx^a$ is included in a
p-form.  \\
K\" ahler found how the Dirac equation could be written as an
equation\cite{kah} ( Eq. (26.6) in the K\" ahler's paper)
\begin{equation}
-i \delta u = m \vee u, \quad {\rm with} \quad \delta u =
\sum_{i=1}^3 dx^i \vee \frac{\partial u}{\partial 
x^i} - dt \vee \frac{\partial u}{\partial t}.
\label{dk1}
\end{equation}
with $u$ defined in Eq.(\ref{form}). The symbol $\delta$ denotes
the inner differentiation, $ a \in \{ 0,1,2,3\}$ and $m$ means
the electron mass.
\\
For a free massless particle living in a d dimensional
space-time  Eq.(\ref{dk1}) can be rewritten in the form
\begin{equation}
dx^a \vee p_a \;\;u = 0, \quad a = 0,1,2,3,5,...,d.
\label{vee}
\end{equation}
The wave function describing the state of
the spin one half particle is packed into the exterior algebra 
function $u$.

\section{ Parallelism between the two approaches.}

We demonstrate the parallelism between the K\" ahler\cite{kah}
and the one of us\cite{norma1}  approach in 
steps, first paying attention on spin $\frac{1}{2}$ only, as K\"
ahler did. Using simple and transparent deffinitions of the exterior
and interior product in Grassmann space, we generalize the K\"
ahler approach first by defining the two kinds of $\delta$
(Eq.(\ref{dk1})) 
operators on the space of p-forms and accordingly three kinds of
the generators of the Lorentz transformations, two of the
spinorial and one of the vectorial character. We try to put
clearly 
forward how the spinorial degrees of freedom emerge out of
vector objects like the 1-forms or $\theta^a$'s. We then
generalize the p-forms to describe not only spins but also
charges of spin $ \frac{1}{2}$ and spin 0 and 1 objects,
unifying also in the space of forms spins and charges. 

\subsection{  Dirac-K\" ahler equation and  Dirac equation in
Grassmann  space for massless particles.}

We present here, side by side, the operators in the space of
differential forms and in Grassmann space: the ''exterior''
product 
\begin{equation}
dx^a \; \wedge dx^b \; \wedge \cdots , \quad \theta^a \theta^b
\cdots, 
\end{equation}
the operator of ''differentiation''
\begin{equation}
-i\; e^a, \quad p^{\theta a} = -i\; \overrightarrow{\partial}^a
= -i \frac{\overrightarrow{\partial}}{\partial \theta_a}, 
\label{a}
\end{equation}
and the two superpositions  
\begin{eqnarray}
dx^a \;\tilde{\vee}: = dx^a \wedge + \; e^a,  \quad
\tilde{a}^a:= i\;(p^{\theta a} - i \theta^a),
\nonumber
\\
dx^a \;\tilde{\tilde{\vee}}: = i\;( dx^a \wedge - \; e^a),  \quad
\tilde{\tilde{a}}{ }^a:= -(p^{\theta a} + i \theta^a).
\label{att}
\end{eqnarray}
The superposition with the sign $\tilde{ }$ is the one used by
K\"ahler (Eqs.(\ref{dk1}).
\\
One easily finds (see Eqs.(\ref{eq6},\ref{eq7})) the commutation
relations, understood in the generalized way of Eq.(\ref{gc})
\begin{eqnarray}
\{ dx^a \; \tilde{\vee},\; dx^b \; \tilde{\vee} \} = 2
\eta^{ab}, \qquad 
\{ \tilde{a}^a, \tilde{a}^b \}  = 2 \eta^{ab},
\nonumber
\\
\{ dx^a \; \tilde{\tilde{\vee}},\; dx^b \;\tilde{\tilde{\vee}} \} = 2
\eta^{ab}, \qquad 
\{ \tilde{\tilde{a}}{ }^a, \tilde{\tilde{a}}{ }^b \} = 2
\eta^{ab}. 
\end{eqnarray}
Since $ \{ e^a , dx^b \; \wedge \} = \eta^{ab}$ and $ \{e^a , e^b
 \} = 0 = \{dx^a \; \wedge, dx^b \; \wedge
 \}$, while $\{ -i p^{\theta a}, 
\theta^b \} = \eta ^{ab}$ and $\{ p^{\theta a},  p^{\theta b} \}
= 0 = \{ \theta^a, \theta^b \}$, it is obvious that $ e^a 
 $ plays in the p-form formalism the role of the
derivative with respect to a differential $1-$form, similarly as
$ip^{\theta a}$ does with respect to a Grassmann coordinate. 
\\
We find for both approaches the  Dirac-like equations:
\begin{eqnarray}
dx^a\; \tilde{\vee}\; p_a\; u = 0,   \quad
\tilde{a}^a\; p_a\; \Phi(\theta^a) = 0,
\nonumber
\\
dx^a\; \tilde{\tilde{\vee}}\; p_a\; u = 0,  \quad \quad
\tilde{\tilde{a}}{ }^a\; p_a\; \Phi(\theta^a) = 0. 
\label{dkt}
\end{eqnarray}

\noindent 
Taking into account the above definitions it follows that
\begin{equation}
dx^a \; \tilde{ \vee}\; p_a\;\; dx^b\; \tilde{\vee}\;
p_b \;u = p^a\; p_a \; u = 0, 
\qquad \tilde{a}^a\; p_a \;\;
\tilde{a}^b\; p_b\; \Phi(\theta^b) = p^a \; p_a\; \Phi(\theta^b) = 0.
\end{equation}
We see that either $dx^a \; \tilde{ \vee}\; p_a \; u = 0\; $ or
$\; dx^a\; \tilde{\tilde{ \vee}} \; p_a\; u = 0 ,$ similarly as either
$\tilde{a}^a \;p_a \;\Phi(\theta^a) = 0\;
$ or $\; \tilde{\tilde{a}}{ }^a \; p_a \; \Phi(\theta^a) = 0 $
can represent the Dirac-like equation.
\\
Both, $dx^a \; \tilde{\vee}$ and $dx^a \; \tilde{\tilde{\vee}} $
define the algebra of the $\gamma^a$ matrices 
and so  do both $\tilde{a}^a $ and $ \tilde{\tilde{a}}{
}^a$. One would thus be tempted to identify
\begin{equation}
\gamma_{ \hbox{naive} }^a := dx^a \; \tilde{\vee},\qquad \hbox{ or } 
\qquad \gamma_{ \hbox{naive} }^a :=  \tilde{a}^a.
\label{naive}
\end{equation}
But there is a large freedom in defining what to identify with
the gamma-matrices, because except when using $\gamma^0$ as
a parity operation, one has an even number of gamma matrices 
occuring in the physical applications such as construction of 
currents $\bar{\psi}\gamma^a\psi$ or for the Lorentz generators 
on spinors $\frac{-i}{4}\; [\gamma^a,\gamma^b]$. Then all the
gamma matrices can be multiplied by some factor provided it does 
disturb neither their algebra nor their even products.
This freedom might be used to solve, what seems a problem:
\\
Having an odd Grassmann character,
neither $\tilde{a}^a$ nor
$\tilde{\tilde{a}}{ }^a$ and similarly neither $dx^a\;
\tilde{\vee}$ nor $ dx^a\; \tilde{\tilde{\vee}}$ should be
recognized as the Dirac 
$\gamma^a$ operators, since they would change, when operating on
polynomials of $\theta^a$ or on superpositions of p-form 
objects of an odd Grassmann
character to objects of an even Grassmann character. One
would, however, expect - since  Grassmann odd fields 
second quantize to fermions, while Grassmann even fields 
second quantize to bosons -  that the $\gamma^a$ operators do not
change the Grassmann character of the wave functions so that
the canonical quantization of Grassmann odd fields then
automatically assures the anticommuting relations between the
operators of the fermionic fields.
\\
We may propose that accordingly 
\begin{equation}
{\rm either} \quad \tilde{\gamma}^a: = i \; dx^0 \;
\tilde{\tilde{\vee}} \; dx^a \; \tilde{\vee}, \quad {\rm or} \qquad
\tilde{\gamma}^a = i \; \tilde{\tilde{a}}{ }^0 \; \tilde{a}^a 
\label{eq25}
\end{equation}
are recognized as the Dirac $\gamma^a$ operators operating on
the space of $p$-forms or polynomyals of $\theta^a$'s,
respectively, since they both 
have an even Grassmann character and they both fulfil the
Clifford algebra $
\{ \tilde{\gamma}^a, \tilde{\gamma}^b \} = 2 \eta^{ab}.$
( The role of $\tilde{ }$ and $\tilde{\tilde{ }}$ can in either
the K\" ahler case or the case of polynomials in Grassmann space, be
exchanged. )
\\
The two definitions of  gamma-matrices  ((\ref{eq25}),
(\ref{naive})) make only a difference when $\gamma^0$-matrix is
used alone. This $\gamma^0$-matrix has to 
simulate the parity reflection which is 
\begin{equation}
{\rm either} \quad \vec{dx}\rightarrow -\vec{dx}, \quad {\rm or}
\quad 
\vec{\theta}\rightarrow - \vec{\theta}.
\label{parity}
\end{equation}
The ''ugly'' gamma-matrix identifications (\ref{eq25}) indeed
perform this operation. 
\\
K\" ahler did not connect eveness and oddness of the forms with
the statistics. He used the ''naive'' gamma-matrix
identifications (\ref{naive}).
The same can be said for
the  Becher-Joos (\cite{bj}) paper.  

\subsection{ Generators of  Lorentz transformations.}

We are presenting the generators of the Lorentz
transformations of spinors for both approaches
\begin{equation}
M^{ab} = L^{ab} + {\cal S}^{ab}, \qquad L^{ab} = x^a p^b - x^b p^a.
\end{equation}
The two approaches  differ in the definition of the generators
of the Lorentz transformations in the internal space 
${\cal S}^{ab}$.
While K\" ahler suggested the definition for spin $\frac{1}{2}$
particles 
\begin{equation}
{\cal S}^{ab} = dx^a \wedge dx^b, \quad {\cal S}^{ab} u = \frac{1}{2}
( (dx^a \wedge dx^b) \vee u - u \vee ( dx^a \wedge dx^b )),
\label{sk}
\end{equation} 
in the Grassmann case\cite{norma1} the two kinds of the
operators 
${\cal S}^{ab} $ for spinors can be defined, presented in
Eqs.(\ref{vecsp}), with the properties 
\begin{equation}
[\tilde{S}^{ab}, \tilde{a}^c] = i(\eta^{ac} \tilde{a}^b -
\eta^{bc} \tilde{a}^a), \;\;         
[\tilde{\tilde{S}}{ }^{ab}, \tilde{\tilde{a}}{ }^c] =
i(\eta^{ac} \tilde{\tilde{a}}{ }^b -
\eta^{bc} \tilde{\tilde{a}}{ }^a), \;\; 
[\tilde{S}^{ab}, \tilde{\tilde{a}}{ }^c] = 0 =
[\tilde{\tilde{S}}{ }^{ab}, \tilde{a}^c].  
\end{equation}
Following the approach in Grassmann space one can also in the
K\" ahler case define two kinds of the Lorentz generators for
spinors, which (both) simplify  Eq.(\ref{sk})  
\begin{eqnarray}
\tilde{{\cal S}}^{ab} = -\frac{i}{4} [dx^a \; \wedge +
\;e^a, \; dx^b\; \wedge + \;
e^b ] ,
\quad \tilde{\tilde{\cal S}}{ }^{ab} =
\frac{i}{4} [dx^a\; \wedge -\; e^a 
,\; dx^b \; \wedge - \;e^b ],
\nonumber
\\
\tilde{{\cal S}}^{ab} = -\frac{i}{4} [\tilde{\gamma}^a,
\tilde{\gamma}^b].\qquad \qquad \qquad \qquad
\label{eq31}
\end{eqnarray}
The above definition enables us to define also in the K\" ahler
case 
the generators of the 
Lorentz transformations of the vectorial character 
\begin{equation}
{\cal S}^{ab} = \tilde{S}^{ab} + \tilde{\tilde{S}}{ }^{ab} = -i
( dx^a \wedge e^b  - dx^b \wedge e^a  ), \quad {\cal
S}^{ab} = \tilde{S}^{ab} + \tilde{\tilde{S}}{ }^{ab} = \theta^a 
p^{\theta b} - \theta^b p^{\theta a}.
\label{vecsk}
\end{equation}
 The operator
$ {\cal S} ^{ab} = -i (dx^a \wedge e^b  - dx^b \wedge e^a),  $ 
being applied on differential p-forms, transforms vectors
into vectors.

\subsection{ Four copies of Weyl bi-spinors  in  K\" ahler
or in approach in Grassmann space and  vector representations.}

 In the case of $d = 4$
one may arrange the space of $2^d $ vectors into four copies of 
two Weyl spinors, one left ( $<\tilde{\Gamma}^{(4)}> = - 1,\;
\;\; \Gamma^{(4)} = 
i\frac{(-2i)^2}{4!} \epsilon_{abcd} {\cal S}^{ab}{\cal S}^{cd}$
) and one right ( $< \tilde{\Gamma}^{(4)}> = 1$) handed (we
have made a choice of $\tilde{ }$ ),
in such a way that they are at the same time the eigen vectors
of the operators $\tilde{S}^{12}$ and the
$\tilde{S}^{03}$ and have  
either an odd or an even Grassmann character. These vectors
are in the K\" ahler approach the superpositions of p-forms and
in the one of us\cite{norma1} approach the polynomials 
 of $\theta^m$'s, $m \in (0,1,2,3)$. The two Weyl
vectors of one copy of the Weyl bi-spinors are connected by the
$\tilde{\gamma}^m$ 
(Eq.(\ref{eq25})) operators.
\\
Analysing the irreducible representations of the group $SO(1,3)$
with respect to the generator of the Lorentz transformations of
the vectorial type\cite{norma1} (Eqs.( \ref{vecsk})) one finds
for d = 4 two scalars ( a scalar and 
a pseudo scalar), two three vectors (in the $SU(2) \times SU(2)
$ representation of $SO(1,3)$ denoted by $(1,0) $ and $(0,1)$
representation, respectively, with $<\Gamma^{(4)} = \pm 1>$) and
two four vectors.   

\subsection{Generalization to extra dimensions.}

 It has  been
suggested\cite{norma1} that the Lorentz transformations in the
space of $ \theta^a$'s in  $d-4$ dimensions manifest
themselves as generators for charges observable at the end for
the four dimensional particles. Since both the
extra dimension spin degrees of freedom and the ordinary spin
degrees of freedom originate from the $\theta^a$'s or the forms
we have a unification of these internal degrees of freedom.
\\
Let us take as an example the model\cite{norma1} which has
$d=14$ and at first - at the high energy
level - $SO(1,13)$ Lorentz group, but which should be broken 
( in two steps ) to first $SO(1,7)\times SO(6)$ and then to
$SO(1,3)\times SU(3)\times SU(2) $.

\section{ Appearance of spinors.}

One of course  wonders about how it is at all possible 
that  the Dirac equation appears
for a {\em spinor} field  out of models with only scalar,
vector and tensor objects! 
 It  only can be done by  {\em exchanging} the Lorentz 
generators ${\cal S}^{ab}$ by the $\tilde{S}^{ab}$ say ( or the
$\tilde{\tilde{S}}{ }^{ab}$ if we 
choose them instead), see equations (\ref{vecsp}, \ref{eq31}).
This indeed means that one of the two kinds of operators
fulfilling the Clifford 
algebra and anticommuting with the other kind - it has been made a
choice of  $dx^a
\tilde{\tilde{\vee}} $ in the K\" ahler case and 
$\tilde{\tilde{a}}{ }^a $ in the approach of one of us - are put
to zero in the operators of Lorentz transformations; as well as in
all the operators representing physical quantities. The  use
of $dx^0 \tilde{\tilde{\vee}}$ or $ \tilde{\tilde{a}}{ }^0 $ in
the operator $\tilde{\gamma}^0$ is the exception  only used to
simulate the Grassmann even  parity operation $\vec{dx}^a \to -\vec{dx}^a $
and $ \vec{\theta} \to - \vec{\theta}, $ respectively.
\\
In (\cite{norma1}) the $\tilde{\tilde{a}}{ }^a$'s are argued away
on the ground of the action.

\section*{Acknowledgement } This work was supported by Ministry of
Science and Technology of Slovenia. One of the authors (H. B.
Nielsen) would like to thank the funds CHRX-CT94-0621, INTAS
93-3316, INTAS-RFBR 95-0567.



\begin{thebibliography}{99}


\bibitem{kah} E. K\" ahler, {\it Rend. Mat. Ser. V}, {\bf 21}, 452
(1962), 
\bibitem{norma1} N. Manko\v c Bor\v stnik, {\it Phys. Lett.} {\bf B  
292}, 25 (1992); {\it Nuovo Cimento} A {\bf 105}, 1461 (1992);
{\it J. Math. Phys.} {\bf
34}, 3731 (1993); {\it Int. Jour. Mod. Phys.} A {\bf 9}1731
(1994); {\it J.
Math. Phys.} {\bf 36}, 1593 (1995); 
{\it Mod. Phys. Lett.} A {\bf  10}, 587 (1995); hep-th9408002;
hep-th9406083;N. Manko\v c Bor\v stnik, S. Fajfer, {\it Nuovo
Cimento} B
{\bf 112 }, 1637 (1997)
\bibitem{bj} P. Becher, H. Joos, {\it Z. Phys. C - Part. and Fields}
{\bf 15}, 343 (1982),
\bibitem{nn} H. B. Nielsen, M. Ninomija, {\it Phys. Lett.} {\bf
B 105}, 219 (1981), {\it Nucl. Phys.  } {\bf B 185}, 20 (1981),
\bibitem{Bled5c11} H. B. Nielsen, N. Manko\v c Bor\v stnik,
{\it Proceedings to the international workshop on What comes beyond
the Standard model, Bled,
Slovenia, 29 June-9 July 1998,} Ed. by N. Manko\v c Bor\v stnik,
H. B. Nielsen, C. Froggatt, DMFA Zalo\v zni\v stvo 1998, p. 68,
\end{thebibliography}
\end{document}